\documentclass[]{emulateapj}

\begin{document}

\title{Evidence of thin helium envelopes in PG1159 stars}

\author{L. G. Althaus\altaffilmark{1}}
\author{A. H. C\'orsico\altaffilmark{1}}
\author{M. M. Miller Bertolami\altaffilmark{1}}
\affil{Facultad de Ciencias Astron\'omicas y Geof\'{\i}sicas,
       Universidad  Nacional de  La Plata,  
       Paseo del  Bosque s/n,
       (1900) La Plata, 
       Argentina}
\email{althaus@fcaglp.unlp.edu.ar} 
\email{acorsico@fcaglp.unlp.edu.ar}
\email{mmiller@fcaglp.unlp.edu.ar}
\author{E. Garc\'{\i}a--Berro\altaffilmark{2}}
\affil{Departament de F\'\i sica Aplicada, 
       Escola Polit\`ecnica Superior de Castelldefels,
       Universitat Polit\`ecnica de Catalunya,  
       Av. del Canal Ol\'\i mpic, s/n,  
       08860 Castelldefels,  
       Spain}
\email{garcia@fa.upc.edu}
\and
\author{S. O. Kepler}
\affil{Instituto de F\'\i sica,
       Universidade Federal do Rio Grande do Sul,
       91501-970 Porto Alegre, RS, 
       Brazil}
\email{kepler@if.ufrgs.br}
\altaffiltext{1}{Member of the Carrera del Investigador Cient\'{\i}fico y 
                 Tecnol\'ogico, CONICET (IALP), Argentina}
\altaffiltext{2}{Institut d'Estudis Espacials de Catalunya, 
                 c/ Gran Capit\`{a} 2--4, 
                 08034 Barcelona, 
                 Spain}
\begin{abstract}
We present evidence that PG1159 stars could harbour He--rich envelopes
substantially  thinner than  those predicted  by  current evolutionary
models with current estimates of  mass loss, which may be attributable
to an  extensive mass--loss episode during the  born--again AGB phase.
Specifically,  we show that  the models  with thin  He--rich envelopes
predict remarkably large  magnitudes of the rates of  period change of
the trapped and  untrapped modes observed in the pulsating star 
PG 1159$-$035. This is a
consequence of  the much shorter evolutionary timescale  of the models
with thin  He--rich envelopes  during the low--gravity  PG1159 regime.
Our findings are particularly interesting in view of the suggestion of
an  evolutionary  link   between  the  helium--deficient  PG1159  star
H1504+65  and the recently  discovered white  dwarfs with  almost pure
carbon atmospheres.
\end{abstract}


\section{Introduction}

PG1159  are  hot  and luminous  stars  that  belong  to the  class  of
hydrogen--deficient post--asymptotic  giant branch (AGB)  stars. These
stars are  supposed to be the main  progenitors of hydrogen--deficient
white dwarfs, which  account for $\sim15 \%$ of  the whole white dwarf
population (Eisenstein et al.  2006).  Currently, about 40  stars are members
of  the PG1159  spectroscopic family  (Werner \&  Herwig  2006), which
spans  a wide  range of  surface gravities  --- $5.5  \lesssim  \log g
\lesssim 8$ (in cgs units) --- and effective temperatures --- 75,000 K
$\lesssim T_{\rm  eff} \lesssim$ 200,000 K.  PG1159  stars are thought
to be  formed as a  result of a  born--again episode, that is,  a very
late thermal pulse (VLTP) experienced  by a hot white dwarf during its
early  cooling phase  --- see  Sch\"onberner  (1979) and  Iben et  al.
(1983) for early  references --- or  a late thermal pulse  that occurs
during the post--AGB  evolution when H burning is  still active (Bl{\"
o}cker 2001).   During the  VLTP, the helium  flash--driven convection
zone   reaches  the   hydrogen--rich   envelope  of   the  star   and,
consequently, most of the hydrogen  is burnt.  The star is then forced
to evolve rapidly back to the AGB and finally as the central star of a
planetary  nebula at high  $T_{\rm eff}$  values.  A  striking feature
characterizing  PG1159  stars   is  their  peculiar  surface  chemical
composition.   Spectroscopic analyses have  revealed that  most PG1159
stars exhibit helium--,  carbon-- and oxygen--rich surface abundances.
The  typical  surface mass  abundances  of  PG1159  stars are  $X_{\rm
He}\simeq  0.33$, $X_{\rm  C}\simeq 0.5$  and $X_{\rm  O}\simeq 0.17$,
though notable  variations are  found from star  to star  (Dreizler \&
Heber 1998; Werner 2001).  The variety of surface patterns observed in
PG1159  stars  poses indeed  a  challenge  to  the theory  of  stellar
evolution.  In particular, the  appreciable abundance of oxygen in the
atmospheres of  these stars has been successfully  explained by Herwig
et  al.  (1999)  on  the  basis of  evolutionary  calculations of  the
born--again scenario that incorporate convective overshoot.

Interest in PG1159 stars is also  motivated by the fact that more than
ten  of them,  the so--called  DOVs, exhibit  multiperiodic luminosity
variations, attributable to global non--radial $g$--mode pulsations, a
fact that has attracted much attention --- see, for instance, Gautschy
et  al.   2005;  C\'orsico   et  al.  2006.   Recently,  considerable
observational effort has  been devoted to the study  of some pulsating
PG1159 stars.   In particular, Vauclair et al.   (2002) have presented
asteroseismological results for  {\mbox{RX J2117.1+3412}} on the basis
of a multisite photometric campaign.  More recently, Fu et al.  (2007)
have detected  a total of  23 frequencies in {\mbox{PG  0122+200}} and
Costa  et  al.   (2008) have  enlarged  to  198  the total  number  of
pulsation  modes in {\mbox{PG  1159$-$035}} ---  the prototype  of the
class --- making it the star with the largest number of modes detected
besides the Sun.

Parallel to these observational breakthroughs, substantial progress in
the theoretical modeling of PG1159  stars has been possible (Herwig et
al. 1999;  Althaus et al.  2005;  Lawlor \& MacDonald  2006).  In this
sense, the full set of  PG1159 evolutionary models developed by Miller
Bertolami \&  Althaus (2006)  has proved to  be valuable  for deriving
structural parameters  of pulsating PG1159 on the  basis of individual
period  fits  ---  see  C\'orsico  et al.  (2007a)  and  C\'orsico  et
al.  (2007b), respectively, for  an application  to the  hot pulsating
star {\mbox{RX J2117.1+3412}} and to  the coolest member of the class,
{\mbox{PG 0122+200}}.  These evolutionary  models are derived from the
complete  evolutionary  history  of  progenitor stars  with  different
stellar  masses   and  an  elaborate  treatment  of   the  mixing  and
extra--mixing  processes during the  violent hydrogen  burning episode
that occurs during the born--again phase.  The success of these models
at  explaining both the  spread in  the surface  chemical compositions
observed in PG1159 stars and the location of the DOV instability strip
in  the  $g-T_{\rm  eff}$  plane  (C\'orsico  et  al.   2006)  renders
reliability to  the inferences drawn from  individual pulsating PG1159
stars.

However, except  for {\mbox{PG 0122+200}}, the  stellar masses derived
from  individual  period  fits  differ considerably  from  the  masses
obtained  using  spectroscopic  data  --- although  the  spectroscopic
uncertainty ($\Delta \log g\simeq  0.5$) is large.  This is noteworthy
in   the   case   of   {\mbox{RX   J2117.1+3412}},   for   which   the
asteroseismological mass is about  25\% smaller than the spectroscopic
value (C\'orsico  et al. 2007a). This discrepancy  could be indicative
of  a missing  piece of  physics in  the existing  PG1159 evolutionary
models, a suspicion  that becomes more solid in  light of the detailed
seismological  study  of  the  pulsating star  {\mbox{PG  1159$-$035}}
recently presented by C\'orsico  et al. (2008).  Indeed, the best--fit
model derived  by these authors for {\mbox{PG  1159$-$035}} is located
outside the  predicted dipole  instablity strip.  Also,  the best--fit
model fails  to predict  a mixture of  positive and negative  rates of
change of  the periods ($\dot \Pi$)  as recently reported  by Costa \&
Kepler (2008)  in {\mbox{PG  1159$-$035}}.  But most  importantly, the
best--fit  model and the  evolutionary models  of Miller  Bertolami \&
Althaus (2006) --- and of other authors as well --- do not predict the
large magnitude  of the rates  of period change observed  in {\mbox{PG
1159$-$035}} (Costa  et al. 1999).   This is a well  known shortcoming
--- see Kawaler \&  Bradley (1994) and C\'orsico et  al. (2008) --- of
the  standard  evolutionary  models  of  PG1159  stars  which  predict
theoretical  $\dot \Pi$ values  about one  order of  magnitude smaller
than those observationally derived.

We   present  here  solid   evidence  that   PG1159  stars   could  be
characterized  by  subtantially  thinner helium--rich  envelopes  than
traditionally accepted  from the standard theory of  the formation and
evolution of  PG1159 stars (Herwig  1999; Miller Bertolami  \& Althaus
2006) that  predicts that  post--born again  remnants are  expected to
retain thick He envelopes. For instance,  the mass of  the He
envelope  for the $0.55\,  M_{\sun}$ sequence  of Miller  Bertolami \&
Althaus (2006) after emerging  from the born--again episode amounts to
$0.03\, M_{\sun}$.   The possibility  that PG1159 stars  could harbour
thinner  He  envelopes  is  sustained  by  the  fact  that  V4334  Sgr
(Sakurai's  object), a star  emerging from  a born--again  episode, is
displaying very strong mass--loss  episodes as it is rapidly reheating
(van Hoof et al.  2007).   In this letter we compute full evolutionary
sequences for  post--born again stars  under the assumption  that such
mass--loss episodes  reduce the mass of the  He envelope considerably.
We  find that  PG1159 evolutionary  sequences with  thin  He envelopes
evolve much faster  than the standard ones.  On  the basis of detailed
pulsation  calculations,  we find  that  PG1159  models  with thin  He
envelopes  predict  rates of  period  change  which  are in  excellent
agreement  with  observations  in  the case  of  {\mbox{PG1159$-$035}}
(Costa  \& Kepler  2008),  thus solving  the longstanding  discrepancy
between the  observed and  theoretical rate of  period change  in this
star.

\section{Evolutionary sequences}

For the  purpose of this  letter we have calculated  full evolutionary
sequences appropriate for PG1159 stars that were evolved directly from
the born--again phase.  The calculations  have been done with the {\tt
LPCODE} evolutionary code that computes the formation and evolution of
white dwarf stars through late thermal pulses (Althaus et al. 2005) on
the basis of a detailed description of the physical processes involved
in the  calculation of the violent hydrogen--burning  event during the
born--again    stage,   particularly   diffusive    overshooting   and
non--instantaneous  mixing. This code  has been  employed to  derive a
full  set   of  PG1159  evolutionary  sequences   starting  from  ZAMS
progenitors with initial  mass ranging from 1.0 to  $3.75 \, M_{\sun}$
(Miller Bertolami  \& Althaus 2006).   All of the sequences  have been
followed from the ZAMS through the thermally--pulsing phase on the AGB
and finally to  the born--again stage where the  remaining hydrogen of
the remnants is burnt.

As  recently discussed  in  Miller Bertolami  \&  Althaus (2007a)  the
born--again models  of Miller  Bertolami \& Althaus  (2006) hint  at a
consistent  picture with  the observed  behavior of  Sakurai's object,
particularly  its extremely fast  post--VLTP evolution.   In addition,
very strong mass--loss episodes during the Sakurai stage appears to be
a  necessary  ingredient  in  the  evolutionary  calculations  (Miller
Bertolami \& Althaus 2007b)  to bring the evolutionary timescales into
agreement with  the short reheating timescale of  the Sakurai's object
after its maximum brightness reported by Van Hoof (2007).  Remarkably,
strong mass--loss episodes have recently  been reported by Van Hoof et
al.  (2007), who  suggested that the mass ejected  by Sakurai's object
in the VLTP could be as high as $10^{-2}\, M_{\sun}$.  In this letter,
we will concentrate on the  consequences of such mass--loss events for
the subsequent evolution of the born--again remnants during the PG1159
stage.  To  this end, we  have considered a rather  extreme situation,
but  still reasonably,  consistent with  the  results of  Van Hoof  et
al. (2007), in which a large  fraction of the remaining He envelope is
lost   as   a   consequence    of   mass--loss   episodes   once   the
hydrogen--deficient star  is back to  the AGB.  Specifically,  we have
concentrated on the $0.556  \, M_{\sun}$ post--VLTP sequence of Miller
Bertolami \& Althaus (2007a), which after the VLTP is characterized by
a He envelope  of $0.032 \, M_{\sun}$.  We assume  that the star loses
$\sim  0.02   \,  M_{\sun}$  as  a  result   of  mass--loss  episodes.
Hereinafter,  this   sequence  will  be   referred  to  as   the  thin
He--envelope sequence. This allows the study PG1159 evolutionary model
sequences with  structurally different envelopes  from those predicted
by  the standard  theory  of stellar  evolution  for post--born  again
stars.

\section{Results and implications}

The most important consequence of the strong envelope reduction caused
by mass--loss  episodes during  the Sakurai stage  is that  the helium
shell burning becomes  virtually extinct. The remnant is  left with no
available nuclear sources and, as  a result, its evolution through the
low--gravity PG1159  regime proceeds on a much  shorter timescale than
in  the  standard PG1159  sequences  in  which  helium burning  mostly
contributes  to  the surface  luminosity  of  the  models. The  marked
reduction  in the  evolutionary timescales  characterizing  the models
with  thin He envelopes  is expected  to yield  large rates  of period
change  in low--gravity  pulsating PG1159  stars, which  is  a feature
needed  to   explain  the   large  $\dot\Pi$  measured   in  {\mbox{PG
1159$-$035}}  itself.  For  a quantitative  inference of  the possible
impact,  we have  computed  $\ell= 1$  $g$--modes adiabatic  pulsation
periods for  our PG1159 models  using the pulsation code  described in
C\'orsico \& Althaus (2006).   In particular, we concentrate on models
belonging  to   our  sequences  with  thin  and   thick  He  envelopes
characterized by  effective temperatures and  surface gravities around
the spectroscopically--determined values for {\mbox{PG 1159$-$035}}.

\begin{figure}
\begin{center}
\includegraphics[clip,width=0.9\columnwidth]{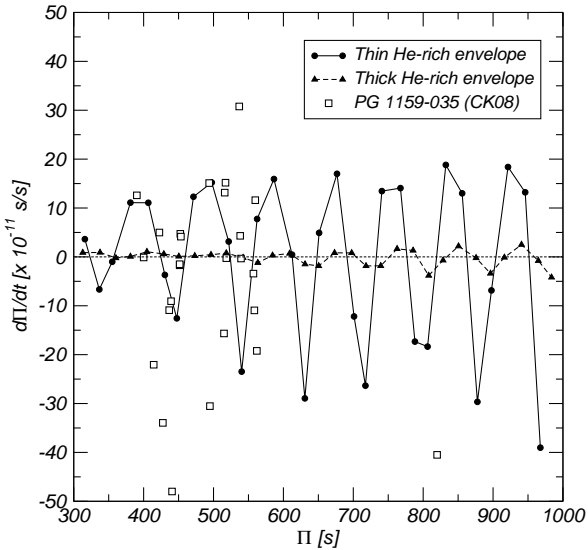}
\caption{The rate of  period change for our PG1159  models.  The solid
         (dashed) line and symbols  correspond to the model characterized
         by a thin  (thick) He envelope. The open  squares display the
         observed  rates of  period change  in  {\mbox{PG 1159$-$035}}
         according to Costa \& Kepler (2008). Note that the model with
         a thin  He envelope predicts the  remarkably large variations
         of the  rates of period  change of the trapped  and untrapped
         modes observed  in this star.}
\label{pdot}
\end{center}
\end{figure}

\begin{figure}
\begin{center}
\includegraphics[clip,width=0.9\columnwidth]{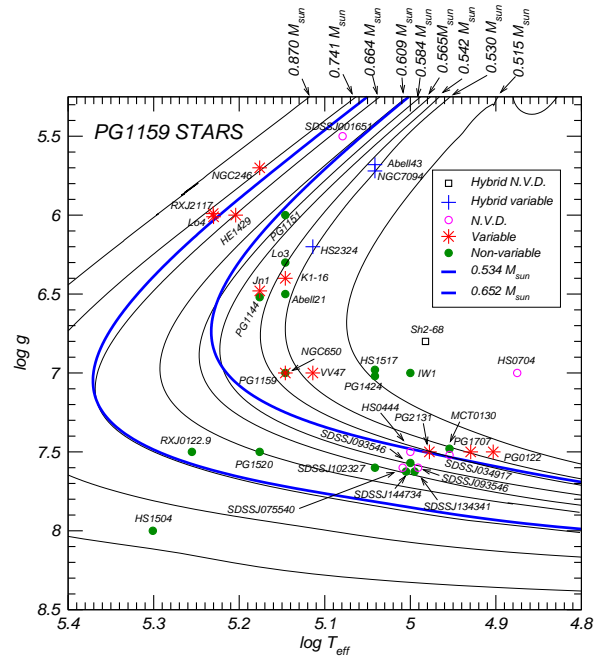}
\caption{The   thick  He--layer   evolutionary  sequences   of  Miller
         Bertolami \& Althaus (2006)  in the $g-T_{\rm eff}$ plane for
         different  stellar masses ---  thin black  lines ---  and the
         0.535  and   $0.652  \,  M_{\sun}$   PG1159  model  sequences
         characterized  by thin He--rich  envelopes described  in this
         paper ---  thick blue lines.  Note that  these sequences have
         substantially  smaller  surface  gravities than  their  thick
         He--rich  counterparts,  particularly  for  the  low--gravity
         regime.   The available  observational data  for  selected PG
         1159 stars are also included. Spectroscopic uncertainties are
         within 0.5  dex for the gravity  and between 10  and 15\% for
         the  effective temperature.  See  the online  version of  the
         journal for a color version of this figure.}
\label{grave}
\end{center}
\end{figure}

In Fig. \ref{pdot}, which summarizes  the main result of this work, we
show the  rate of period change  $\dot{\Pi}$ for periods  in the range
$300  \leq \Pi  \leq 1000$  s.   The results  corresponding to  PG1159
models with thin (thick) He envelopes are shown with solid (dashed) lines
and symbols.  The model with thin He envelope is characterized by
a stellar mass of $0.534 \, M_{\sun}$, a He envelope of $0.012 \, M_{\sun}$, 
$T_{\rm   eff} \sim 160,000$ K, log L/$L_{\sun} \sim 2.88$, and log
$g \sim 7.1$. In both models,  modes trapped in the outer envelope are
characterized  by  negative  rates  of  period  change.   This  figure
demonstrates that the  values of $\dot{\Pi}$ of the  model with a thin
He envelope  far exceed those predicted  by the model with  a thick He
envelope.   This is  a consequence  of the  much  shorter evolutionary
timescale of the model with a thin He envelope during the low--gravity
PG1159 regime.  Indeed, for  untrapped modes ($\dot{\Pi}>0$) the rates
can be up to an order of  magnitude larger in the case in which a thin
He   envelope  is  adopted,   and  even   larger  for   trapped  modes
($\dot{\Pi}<0$).   On the  contrary,  for the  regime  of the  evolved
pulsating PG1159  stars of high gravity such  as {\mbox{PG 0122+200}},
already  on the  hot white  dwarf  cooling track,  where helium  shell
burning is no longer the luminosity source in the standard models, the
magnitude of  the period  changes becomes much  less sensitive  to the
thickness  of the  helium envelope,  as  noted earlier  by Kawaler  \&
Bradley  (1994).  In  Fig.  \ref{pdot}  we also  include  the observed
rates  of period  change in  {\mbox{PG 1159$-$035}}  (Costa  \& Kepler
2008),  which are  displayed  as open  squares.   Note that  {\mbox{PG
1159$-$035}}  exhibits a  mixture of  positive and  negative  rates of
large amplitude (up to $\approx 40 \times 10^{-11}$ s/s).  The general
agreement between the $\dot{\Pi}$ values predicted by the model with a
thin He--rich envelope and those observed in {\mbox{PG 1159$-$035}} is
remarkable.   Note  in  particular  that  the rate  of  period  change
[$\dot{\Pi}= (+15.17 \pm  0.045) \times 10^{-11}$ s/s] of  the 517.1 s
mode --- the central peak of one of the {\mbox{PG 1159$-$035}} highest
amplitude  modes ---  is  well reproduced  by  the model  with a  thin
He--rich  envelope, in  sharp  contrast with  the  predictions of  the
standard  model  with  a  thick  envelope, for  which  the  values  of
$\dot{\Pi}$  remain  smaller  than  $2\times  10^{-11}$  s/s  for  the
relevant range of  periods.  From these results, it  is clear that the
presence of a thin He--rich envelope in {\mbox{PG 1159$-$035}} appears
to    be   a    key   ingredient    to   resolve    the   longstanding
order--of--magnitude discrepancy  between the predictions  of standard
PG1159 evolutionary models and the  observed rates of period change in
{\mbox{PG 1159$-$035}}. We want to mention that the effective temperature of 
our model with thin He envelope is somewhat larger than the spectroscopically 
inferred value for {\mbox{PG 1159$-$035}}. However, this inconsistency does not
affect the main conclusion of this letter about the strong impact of thin
He envelopes on the predicted rate of period changes in this pulsating
PG1159 star. More precise asteroseismological fits to the observed period 
and period changes in {\mbox{PG 1159$-$035}} with sequences with thin He 
envelopes would require the computation of a much finner grid of evolutionary 
sequences, an issue that would carry us beyond the scope of the present letter.

Another  implication of  PG1159 stars  having thin  He--rich envelopes
concerns their  location in the  $g-T_{\rm eff}$ plane.  In  fact, the
evolutionary tracks in this plane are strongly modified if the mass of
the  He--rich  envelope  is  reduced.    This  can  be  seen  in  Fig.
\ref{grave}  where  the  standard  evolutionary  sequences  of  Miller
Bertolami \& Althaus (2006) --- thin lines --- are shown together with
two sequences of masses 0.535 and $0.652\,M_{\sun}$ computed with thin
He--rich envelopes (solid blue  lines), characterized by
envelopes of 0.012 and $ 0.008\, M_{\sun}$,  respectively 
(as compared with the envelope mass of the standard models of 
0.032 and 0.02 $\, M_{\sun}$).  Note that for the low-gravity
regime, tracks characterized by thin He--rich envolpes are hotter than
their  thick He--rich  counterparts.  That  is, at  a  given effective
temperature,  thin He--rich  sequences are  characterized  by markedly
smaller gravities than the  sequences with thick envelopes.  This will
certainly  affect the  spectroscopic mass  determinations.  This  is a
relevant  issue concerning  the hottest  known pulsating  PG1159 star,
{\mbox{RX  J2117.1+3412}}, a  low--gravity PG1159  star for  which the
spectroscopic mass  --- about $0.72 \, M_{\sun}$  (Miller Bertolami \&
Althaus 2006) --- is much  larger than the seismological mass of $0.56
\, M_{\sun}$ (Vauclair et al. 2002; C\'orsico et al. 2007a). Note that
the models with thin  envelopes predict for {\mbox{RX J2117.1+3412}} a
spectroscopic  mass of about  $0.65\, M_{\sun}$,  considerably smaller
than that derived from  thick He-rich models.  Hence, PG1159 sequences
with  thin He--rich envelopes  may help  to alleviate  the discrepancy
between   the   spectroscopic    and   seismic   mass   of   {\mbox{RX
J2117.1+3412}}.

\section{Conclusions}

In this letter we have presented  evidence that PG1159 stars --- or at
least  {\mbox{PG  1159$-$035}}   itself  ---  could  harbour  He--rich
envelopes thinner  than those predicted previously.   The existence of
PG1159 stars with thin He--rich envelopes is in line with the reported
strong mass--loss  episodes in the  Sakurai's object (Van Hoof  et el.
2007), an emerging post--born again star that is expected to evolve to
the PG1159  domain.  Specifically,  we have shown  that if  the PG1159
progenitors  emerge from their  born--again episode  with considerably
reduced  envelopes,  then the  predicted  rates  of  period change  in
low--gravity pulsating PG1159 stars  are profoundly altered to such an
extent  that the  longstanding  discrepancy with  the  rate of  period
change  measured in  {\mbox{PG  1159$-$035}} is  solved.  This  result
opens  the   possibility  that  PG1159  stars  in   general  could  be
characterized  by   thinner  He--rich  envelopes   than  traditionally
accepted on the  basis of the standard theory  of post--AGB evolution.
Pushing our results to their  limits, we are tempted to speculate that
the  helium deficiency observed  in H1504+65,  the most  massive known
PG1159 star, could be the  result of stronger mass--loss episodes than
those observed  in Sakurai's object,  which combined with  the smaller
helium content that remains in massive post--VLTP stars, could lead to
PG1159 remnants  almost devoided of  helium.  The existence  of PG1159
stars with thin helium  envelopes, is particularly interesting in view
of  the   suggestion  of   an  evolutionary  connection   between  the
helium--deficient  PG1159 star  H1504+65 and  the  recently discovered
white dwarf population with carbon atmospheres (Dufour et al. 2007).

\acknowledgments

Part of this work was supported by PIP 6521 grant from CONICET, by MEC
grant AYA05-08013-C03-01,  by the European  Union FEDER funds,  and by
the AGAUR.

\end{document}